\documentclass[aps,a4paper,twocolumn]{revtex4}

\usepackage{graphicx}
\usepackage{amsmath}
\usepackage{amssymb}

\newcommand{\ud}{\,{\mathrm d}}

\begin{document}
\title{Stability of magnetic vortex in soft magnetic nano-sized circular
  cylinder}
\author{Konstantin L. Metlov}
\affiliation{Institute of Physics ASCR, Na Slovance 2, Prague 8, CZ-18200}
\email{metlov@fzu.cz}
\author{Konstantin Yu. Guslienko}
\affiliation{Korea Institute for Advanced Study, 207-43 Cheongryangri-dong, Dongdaemun-gu, Seoul 130-012 Korea}
\date{\today}
\begin{abstract}
  Stability of magnetic vortex with respect to displacement of its
  center in a nano-scale circular cylinder made of soft ferromagnetic
  material is studied theoretically.  The mode of vortex displacement
  producing no magnetic charges on the cylinder side is proposed and
  the corresponding absolute single-domain radius of the cylinder is
  calculated as a function of its thickness and the exchange length of
  the material. In cylinders with the radii less than the
  single-domain radius the vortex state is unstable and is absolutely
  prohibited (except if pinned by material imperfections), so that the
  distribution of the magnetization vector in such cylinders in no
  applied magnetic field is uniform (or quasi-uniform). The phase
  diagram of nano-scale cylinders including the stability line and the
  metastability region obtained here is presented.
\end{abstract}
\maketitle

The small magnetic nano-scale cylinders made of soft magnetic
materials recently gained attention due to the progress in fabrication
and observation techniques and also because of their possible
applications in magnetic random access memory (MRAM) devices. Within
such a cylinders of circular shape in a certain range of sizes the
magnetic vortices are frequently observed (see e.g.
\cite{FGWC98,SOHSO00,SHZ00,PSN00}). For applications, which usually
try to avoid vortex formation (such as MRAM cells, \cite{GSSJTG00}),
it is important to know the sizes of the cylinder where the vortices
do not form.

In thin ferromagnetic cylinders with the thickness $L$ of the order of
a few $L_E$ (where $L_E=\sqrt{C/M_S^2}$ is the exchange length, $C$ is
the exchange constant of the material, $M_S$ is the saturation
magnetization) distribution of the magnetization vector can be assumed
uniform along the cylinder axis. Then, there are two characteristic
sizes of the cylinder important for the presence of the vortex state
versus the uniformly magnetized one in zero applied magnetic field.
The first is the single-domain radius $R_{EQ}$, which is defined as a
radius of the cylinder (at a given thickness) in which the energies of
the uniformly magnetized state and the state with the vortex are the
same \cite{UP93}. In cylinders with radii $R<R_{EQ}$ the uniformly
magnetized state has a lower energy than the vortex state. However,
the metastable vortices still may be present in cylinders with radii
below $R_{EQ}$. There is another characteristic radius, the absolute
single domain radius $R_S$ of the cylinder, which is obtained from the
requirement that the vortex is {\em unstable} (therefore is absolutely
prohibited) in cylinders with $R<R_S$.

The rigorous calculation of $R_S$ requires evaluating the second
variation of the energy functional including the long-range dipolar
interactions, which is currently beyond possibilities of analytical
methods. The other way to estimate the stability radius is to assume
the precise way (mode) the vortex loses its stability and then to
calculate the stability radius with respect to that process. There is
an infinite set of possible candidate modes. Provided calculation of
the energies is rigorous, the result for a particular candidate is a
lower bound for the stability radius. The ``true'' $R_S$ is the
highest of all the lower bounds for all possible modes.  This is also
easy to understand from a thought experiment of having a cylinder with
vortex and slowly shrinking its radius, as soon as the largest
absolute single domain radius is reached the vortex loses stability
according to that particular mode.

The first estimate of the absolute single-domain radius $R_S$ was
obtained for the circular cylinder by Usov and coworkers \cite{UP94}
for the mode corresponding to the uniform translation of the vortex.
Such a translation is accompanied by formation of magnetic charges
on the cylinder sides.

In {\em small} and {\em flat} ferromagnetic cylinders of isotropic
(ideally soft) material the following hierarchy of energies is
present\cite{M01_solitons2} (by decreasing their importance for
minimization): exchange energy, energy of magnetic charges of the
cylinder faces (because the cylinder is flat), energy of magnetic
charges on cylinder sides, energy of volume magnetic charges. The
energy of volume charges is the least important one, because of a
general amplification of the effects associated with surface compared
to the ones associated with volume, as the size of the particle
decreases. Thus, it may be expected (and further calculations in this
work support this conjecture) that a mode with non-uniform vortex
deformation having no side magnetic charges (but having volume ones)
may be more favorable for the vortex stability loss.

\begin{figure}[tbp]
  \begin{center}
    \includegraphics[scale=0.7]{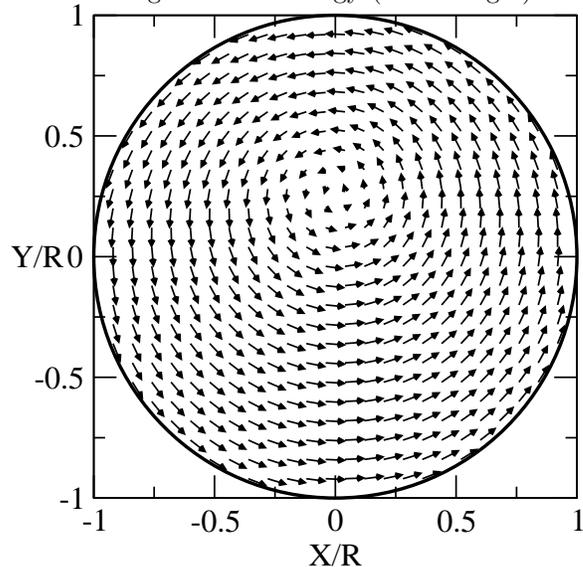}
    \caption{The distribution of the magnetization vector
      (\ref{eq:structure_nochrg}) corresponding to a displaced vortex
      with $c=1/2$, $a=1$. Only projection of the magnetization to the
      cylinder face is shown, shorter arrows mean non-zero
      out-of-plane magnetization component.}
    \label{fig:structure}
  \end{center}
\end{figure}

To find such a mode is not an easy task because once the vortex is
deformed the problem loses the cylindrical symmetry and becomes
two-dimensional (leading to a system of partial differential integral
equations) instead of one-dimensional (described by a single ordinary
differential equation) if the axial symmetry is present. It is
possible to write the solution for the displaced magnetic vortex with
no side magnetic charges and very small face charges at the vortex
center exactly minimizing the exchange energy functional in the
following form \cite{M01_solitons2}
\begin{eqnarray}
  f(z) & = & \frac{1}{c}
  \left(
    \imath z + \frac{1}{2}
    \left(
      a - \overline{a} z^2
    \right)
  \right), \label{eq:structure_nochrg} \\
  w(z,\overline{z}) & = & 
  \left\{ 
    \begin{array}{ll}
      f(z) & |f(z)| \leq 1 \\
      f(z)/\sqrt{f(z) {\overline f}({\overline z})} & |f(z)|>1
    \end{array} 
  \right. , \label{eq:sol_SM}
\end{eqnarray}
where $z=X+\imath Y$, $\imath=\sqrt{-1}$, $X$ and $Y$ are Cartesian
coordinates in the cylinder plane, so that the axis $Z$ is parallel to
the cylinder axis, the complex parameter $a$ describes the vortex
displacement (for $a=0$ the vortex is centered), the real parameter
$c$ is related to the vortex radius, bar over a variable denotes the
complex conjugation, the components of the magnetization vector
$\vec{M}(X,Y)$ are given by $M_X+\imath M_Y = M_S 2 w/(1+w
\overline{w})$ and $M_Z=M_S(1-w \overline{w})/(1+w \overline{w})$. An
example of configuration (\ref{eq:structure_nochrg}) is shown in
Fig.~\ref{fig:structure}. In the case of no applied field (the case
when the field is non-zero was considered elsewhere \cite{GM01}) the
phase of the complex number $a$ is not important, this parameter will
be considered real in the rest of this work. The expression
(\ref{eq:structure_nochrg}) is not arbitrary, it follows from the
analysis performed in \cite{M01_solitons2}, that it is {\em the only
  way} to displace the vortex center so that it's structure keeps
minimizing the exchange energy functional exactly and has no surface
magnetic charges on the cylinder side. The magnetization distribution
(\ref{eq:structure_nochrg}) has surface magnetic charges on the faces
of the cylinder localized near the vortex center, which is a
topological singularity \cite{M01_solitons2}. Unlike side charges, the
charges on the faces can not be avoided if the cylinder is in a state of
non-uniform magnetization (topologically charged state), such as
vortex.

The values of the parameters $a$ and $c$ need to be found from
minimization of the magnetostatic energy for the configuration
(\ref{eq:structure_nochrg}). It is clear that in absence of magnetic
field, the equilibrium $a=0$ due to symmetry, in this case
(\ref{eq:structure_nochrg}) reduces to the structure of an axially
symmetric centered vortex exactly coinciding with the one studied by
Usov\cite{UP93}. The parameter $c=R_V/R$, where $R_V$ is the
equilibrium vortex radius calculated in \cite{UP93}.

The stability radius can be found from the condition
\begin{equation}
  \label{eq:stabrad_cond}
  \left.
    \frac{\partial^2}{\partial a^2} 
    \left(
      E_{EX}(a) + E_{MS}(a)
    \right)
  \right|_{a=0}
    =0,
\end{equation}
which means that the energy minimum on the variable $a$ turns to
maximum. In this expression $E_{EX}$ and $E_{MS}$ are the total
exchange and magnetostatic energy of the cylinder respectively.

The normalized exchange energy of the particle with magnetic
configuration (\ref{eq:structure_nochrg}) is
\begin{equation}
  \label{eq:structure_nochrg_meron_enr}
  e_{EX}=\frac{E_{EX}}{4\pi M_S^2 V}=
  \frac{L_E^2}{4\pi R^2} (2 - \log \frac{2 c}{1+\sqrt{1-a^2}}),
\end{equation}
where $V=L \pi R^2$ is the particle volume.

\begin{figure}[tbp]
  \begin{center}
    \includegraphics[scale=0.45]{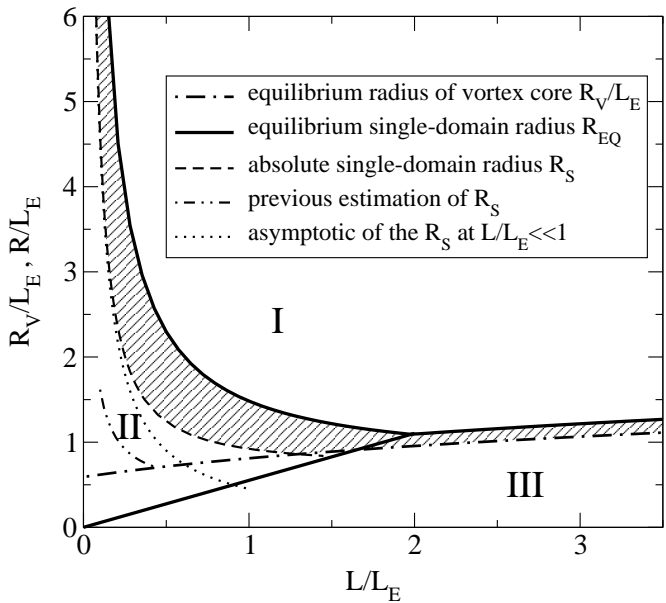}
    \caption{The phase diagram of nano-scale cylinders. The roman
      numbers mark the regions separated by the solid lines
      corresponding to different ground states of the particle
      \cite{UP93,UP94}: I -- vortex state, II -- magnetized uniformly
      in-plane, III -- magnetized uniformly parallel to the cylinder
      axis. The shaded region marks the range of cylinder's
      geometrical parameters where the vortices are metastable, below
      that region no complete vortices can be present. The previous
      estimation of $R_S$ corresponds to the uniform vortex
      displacement \cite{UP94}}
    \label{fig:diagram}
  \end{center}
\end{figure}

Due to the lack of space the change (with $a$) of the vortex core
magnetostatic energy (face charges) will be neglected in this paper.
This approximation is equivalent to neglecting terms of the order
$c^3\ll1$. The density of volume magnetic charges (also neglecting the
fact that inside the vortex core $|f(z)|<1$ they are reduced,
equivalent to throwing out terms of the order $c^5\ll1$) is
\begin{equation}
  \label{eq:divm}
  \vec{\nabla}\cdot\vec{M}(\vec{r}) =
  - \frac{a M_S}{R} \cos \phi + O(a^2),
\end{equation}
in the polar coordinate system $r$, $\phi$ centered at the particle
axis. Their normalized energy is
\begin{equation}
  \label{eq:structure_nochrg_volume_MS}
  e^{\mathrm{vol}}_{\mathrm{MS}} = 
  \frac{E^{\mathrm{vol}}_{\mathrm{MS}}}{4\pi M_S^2 V}= 
  \frac{R a^2}{2L} F_1^V(\frac{L}{R}) + O(a^3)
\end{equation}
where straightforwardly
\begin{widetext}
\begin{equation}
  \label{eq:structure_nochrg_F1V_straightforward}
  F_1^V(x)  =  \frac{1}{(2\pi)^2}
  \int_0^x \int_0^x
  \int_0^1 \int_0^1
  \int_0^{2\pi} \!\!\! \int_0^{2\pi}\!\!\!
  \frac{r_1 r_2 \cos\phi_1 \cos\phi_2 \ud z_1 \ud z_2 \ud r_1
    \ud r_2 \ud \phi_1 \ud \phi_2}
  {\sqrt{(z_1-z_2)^2+r_1^2+r_2^2-2 r_1 r_2 \cos (\phi_1-\phi_2)}},
\end{equation}
\end{widetext}
which can be simplified into
\begin{equation}
  \label{eq:structure_nochrg_F1V_simple}
  F_1^V(x)  =  \int_0^\infty \ud k \frac{x}{k}
  \left( 
    1 - \frac{1-e^{-k x}}{k x}
  \right) 
  \left[
    \int_0^1 \rho J_1(k \rho)\ud\rho
  \right]^2.
\end{equation}
Equation (\ref{eq:stabrad_cond}) was then solved numerically for
$R_S/L_E$ at different $L/L_E$, the result is shown in
Fig.~\ref{fig:diagram}. It is possible to obtain the asymptotic of
this dependence at $L \ll L_E$ analytically by expanding
$F_1^V(x<<1)=k x^2 + O(x^3)$, where $k=0.0882686\ldots$ we get
\begin{equation}
  \label{eq:stab_rad_asympt}
  \frac{R_S}{L_E}=\frac{1}{8\pi k (L/L_E)}, \qquad L \ll L_E.
\end{equation}
This asymptotic was used to verify the results of the numerical
computation and is also plotted in Fig.~\ref{fig:diagram}.

Neglecting the vortex core is not such a bad approximation as it
seems, in fact, it only breaks when $c=R_V/R$ is close to one, which
happens near the intersection of the $R_S$ and $R_V$ lines in
Fig.~\ref{fig:diagram}.  Including the core into consideration is
straightforward but involves more complicated formulas. Such more
detailed calculation will be published in a forthcoming paper.

This work was supported in part by the Grant Agency of the Czech
Republic under projects 202/99/P052, 101/99/1662, the INTAS Grant
31-311 and the Korea Institute for Advanced Study. The authors would
like to thank Ivan Tom{\'a}{\v s} for reading the manuscript and many
valuable discussions.

\end{document}